\documentclass[%
 reprint,
 showpacs,preprintnumbers,
 amsmath,amssymb,
 aps,
prb
]{revtex4-1}
\usepackage{dcolumn}
\usepackage[dvipdfmx]{graphicx}
\usepackage{subfigure,color}
\usepackage{mathrsfs}

\makeatletter
\def\btt#1{\texttt{\@backslashchar#1}}
\DeclareRobustCommand\bblash{\btt{\@backslashchar}} \makeatother

\def\gsim{\lower -0.3ex \hbox{$>$} \kern -0.75em \lower 0.7ex
	\hbox{$\sim$}}
\def\lsim{\lower -0.3ex \hbox{$<$} \kern -0.75em \lower 0.7ex
	\hbox{$\sim$}}

\begin{document}

\title{Spin-dependent Refraction at the Atomic Step of Transition-metal Dichalcogenides}

\author{Tetsuro Habe and Mikito Koshino}
\affiliation{Department of Physics, Tohoku University, Sendai 980-8578, Japan}

\date{\today}

\begin{abstract}
We theoretically propose a novel
spin-dependent electronic transport mechanism
in which the spin-unpolarized electron beam is split
into different directions depending on spins
at an atomic domain boundary in non-magnetic material.
Specifically, we calculate the electronic transmission across a boundary
between monolayer and bilayer of the transition metal dichalcogenide, 
and demonstrate that up-spin and down-spin electrons entering the boundary 
are refracted and collimated to opposite directions.
The phenomenon is attributed to 
the strong spin-orbit interaction, the trigonally-warped Fermi surface, 
and the different crystal symmetries between 
the monolayer and bilayer systems.
The spin-dependent refraction suggests a potential application for a spin splitter, 
which spatially separates up-spin and down-spin electrons 
simply by passing the electric current through the boundary.
\end{abstract}


\maketitle


Spin-dependent electron transport is a key ingredient in spintronics
which exploits the spin degree of freedom for electronic devices.
Particularly, the capability to manipulate spins purely by electric means 
is a desirable property, as it allows the combination with the conventional electronics.
For the electrical control of spins without resorting to magnetic field, 
the spin-orbit interaction plays an essential role.
A variety of spin-dependent transport mechanisms in non-magnetic materials,
such as the spin-Hall effect \cite{hirsch1999spin}
and the spin field effect devices \cite{datta1990electronic},
are derived from the spin-orbit interaction.

\begin{figure}[htbp]
 	\includegraphics[width=70mm]{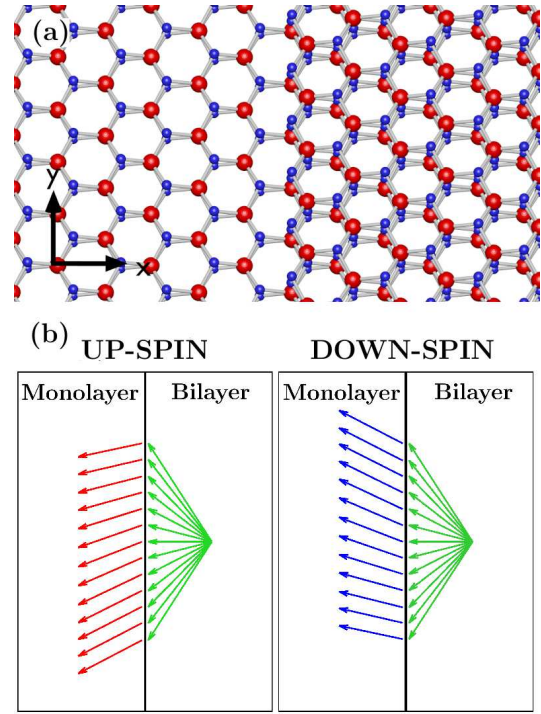}
 	\caption{
 		(a) Atomic structure of the junction between the monolayer and bilayer of TMD.
 		The large (red) and small (blue) spheres represent the transition-metal and chalcogenide atoms, respectively.
 		(b) Electron refraction at the atomic step between monolayer and bilayer of MoTe$_2$
 		for an incident electron from the bilayer side.
 	}\label{fig_schem}
 \end{figure}

In this paper, we theoretically propose a novel spin-dependent transport mechanism,
referred to as spin-dependent refraction in the following,
in which the spin-unpolarized electron beam is split 
into different directions depending on spins
at an atomic domain boundary in non-magnetic material.
Specifically, we consider an atomic step
between monolayer and bilayer of the transition metal dichalcogenide (TMD)
as shown in Figure\ \ref{fig_schem}(a) and demonstrate that
up-spin and down-spin electrons entering from the bilayer side
are refracted and collimated to opposite directions, as 
illustrated in Fig.\ \ref{fig_schem}(b).
The spin-dependent refraction effect can be exploited for a spin splitter, 
which spatially separates up-spin and down-spin electrons 
simply by passing the electric current through the boundary,
just like an optical refraction separating a light beam by the wavelength.

\begin{figure}
\includegraphics[width=90mm]{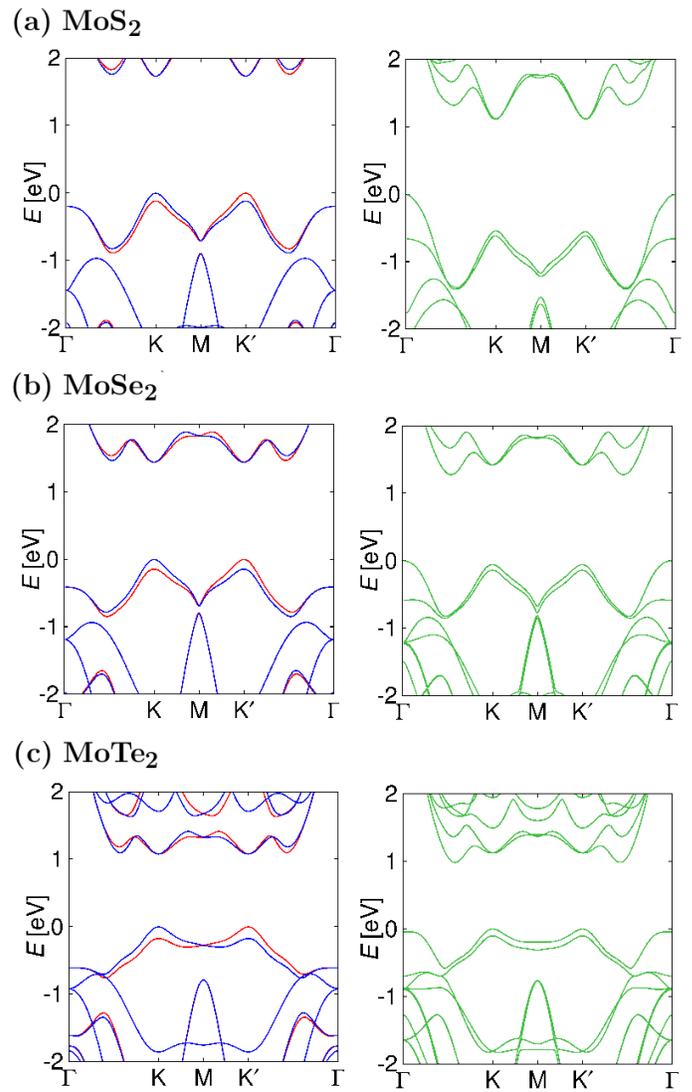}
\caption{Band structures of monolayer (left) and bilayer (right) of 
	(a)MoS$_2$, (b)MoSe$_2$, and (c)MoTe$_2$. 
In monolayer, the up-spin and down-spin states are represented by the red and blue lines, respectively.
In bilayer, the two spin states are degenerate.}\label{fig_band}
\end{figure}

TMD recently attracts a significant attention as a novel family of 
two-dimentional material. \cite{Mak2010,Splendiani2010,Coleman2011,Lee2012,Wang2012,Chhowalla2013,Jin2013}
A hallmark of the electronic structure of the TMD monolayer
is the correlation of the spin and valley degrees of freedom.
Specifically, the valence band maxima located 
at $K$ and $K'$ valleys are spin split
in the opposite direction between the two valleys as shown in Fig.\ \ref{fig_band},
and this is due to the strong spin-orbit coupling of the heavy transition-metal atoms
and also the absence of the inversion symmetry in the lattice structure.
The spin-valley correlated band structure 
leads to characteristic spin-dependent optical properties,
which have been extensively studied in the recent years.
 \cite{Cao2012,Zeng2012,KinFai2012,Shi2013,Molina2013,Suzuki2014,ZhangC2014,Yamamoto2014}.
The properties of TMD atomic layers are also studied
in terms of the electric and spintronic transport\cite{Song2013,Yuan2014,Feng2012,Klinovaja2013,Kormanyos2014,Peng2014}. 
 
Here we propose the spin-dependent refraction 
as a mechanism to manipulate the electronic spins
in TMD without using optics or magnetic field.
We calculate the electronic transmission probability 
across the monolayer-bilayer boundary on several kinds of TMDs
using the tight-binding model based on the first-principle band calculation,
and actually show that the up-spin and down-spin electrons
are refracted different angles at the boundary.
The phenomenon is attributed to the common characteristics of TMDs; the strong spin-orbit interaction, 
the trigonal warped Fermi surface, 
and the difference of the symmetry between the monolayer and bilayer TMDs.
We find that the spin-dependent refraction effect is conspicuous in
TMDs MX$_2$ with M = Mo, W and X = Se, Te,
while it is not observed in the sulphides (X=S),
where the carriers are fully reflected at the atomic step.
Previously, the electron transmission property
was studied for graphene monolayer-bilayer 
junction \cite{Nakanishi2010}, and it was shown that 
the electrons are refracted 
to different angles depending on $K$ and $K'$ valleys.
In TMDs, a valley-dependent transport
immediately leads to a spin-dependent transport, owing to the spin-valley correlation.


We consider an atomic junction between monolayer and bilayer 
of a single sort of TMD, as shown in Fig.\ \ref{fig_schem}.
Here we assume that one layer in the bilayer region
is truncated at the $y$ axis (zigzag direction),
and the other layer continues to the monolayer region.
The bilayer region takes the inversion-symmetric structure
called 2H stacking, which is the most common phase in the bulk TMD.
We also assume that the carrier density in the sample is controlled 
by a single gate electrode underneath and it is homogeneous over the whole system.
To simulate this situation, we appropriately differentiate 
the electrostatic potential of the monolayer and that of the bilayer
to achieve the given carrier density at the common Fermi energy.
To describe the motion of electrons,
we perform the first-principle calculation
using the numerical package of quantum-ESPRESSO\cite{Quantum-espresso},
and obtain the electronic structure 
of infinite TMD monolayer and that of bilayer.
From the obtained first-principle electronic density, 
we compute the maximally-localized Wannier function
using Wannier90 package \cite{Wannier90} and
derive the tight-binding hopping parameters for bulk monolayer and bilayer.
Finally, we apply the tight-binding model to the monolayer-bilayer junction
to calculate the transmission probability with the Green's function method.\cite{Ando1991}
In the first principle calculations, 
we adopt the geometrical parameters for the crystal structure 
in Refs. \onlinecite{Yun2012,Debbichi2014,He2014},
and employ the GGA pseudopotentials, the cutoff energy of the plane-wave basis $150$[eV] 
and the convergence criterion of $10^{-8}$[eV].
The basis of the tight-binding Hamiltonian consists of $d$-orbitals 
on transition-metal atoms and $p$-orbitals on chalcogenide atoms.
The detailed description of the tight-binding Hamiltonian for monoalyer-bilayer
junction is presented in Appendix.




Fig.\ \ref{fig_band} presents the first-principle band structures of the monolayer and bilayer
 of MoS$_2$, MoSe$_2$, and MoTe$_2$.
In the monolayer of each TMD, the valence band is spin-split 
near the peaks at $K$ and $K'$, where the spin is polarized
to the opposite directions in two valleys. \cite{Xiao2012}
In the bilayer, on the other hand, the energy bands are 
completely spin-degenerate due to the inversion symmetry of the lattice structure.
We also notice that the valence band at the $\Gamma$ point  
becomes relatively higher in bilayer than in monolayer.
In MoS$_2$ bilayer, the $\Gamma$ point is higher than 
the $K$ and $K'$ points by about 500meV,
while in MoSe$_2$ bilayer the deference reduces to $100\mathrm{[meV]}$,
and in MoTe$_2$ bilayer, the $\Gamma$ point is slightly lower than the $K$ and $K'$.
The band structures of the tungsten dichalchogenides WX$_2$ (not shown)
exhibit a similar tendency where the $\Gamma$ point energy becomes relatively lower
for heavier chalcogenide atoms. \cite{Yun2012,Kumar2013}

\begin{figure*}
	\includegraphics[width=0.9\textwidth]{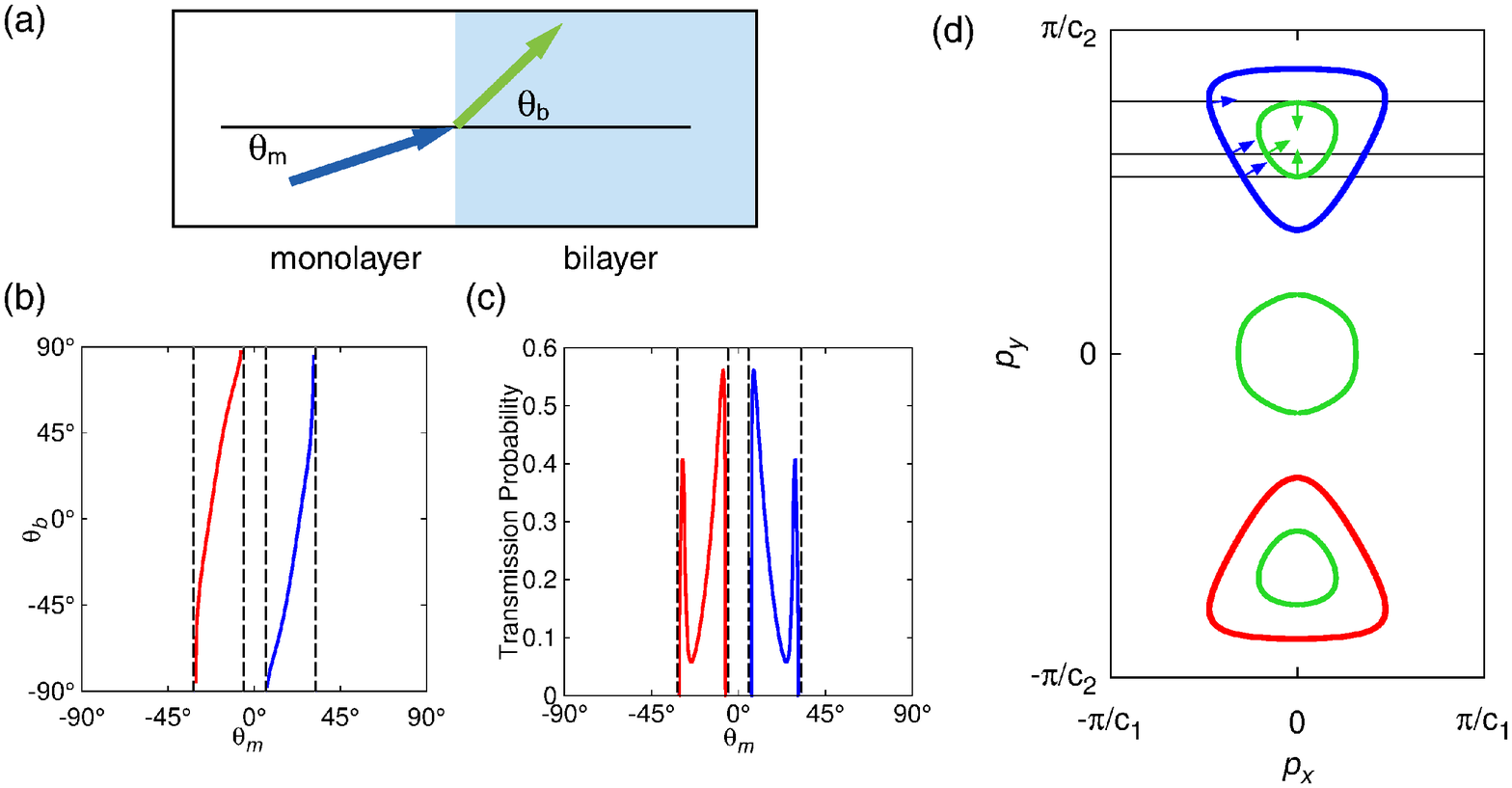}
	\caption{
		Transmission property in the monolayer-bilayer junction of MoTe$_2$
		with the carrier density $n=-7.02\times10^{13}[\mathrm{cm}^{-2}]$.
		(a) Schematic of the refraction process with the 
		definition of the $\theta_m$ and $\theta_b$.
		(b) Relation between the velocity angles $\theta_m$ and $\theta_b$,
		for up-spin (red) and the down-spin (blue) electrons.
		(c) Corresponding transmission probability as a function of $\theta_m$.
		(d) Fermi surfaces of the monolayer and bilayer.  
		The red and blue curves represent up-spin and down-spin states in monolayer, respectively,
		and the green is the spin-degenerate states in bilayer.
		Arrows indicate the velocities of the electrons 
		at some $p_y$'s on the Fermi surface.
	}\label{fig_refract_mote2}
\end{figure*}



Now let us consider a TMD monoalyer-bilayer hybrid system.
In Fig.\ \ref{fig_refract_mote2}, we present some example of 
the calculated spin-dependent electron transmission 
in the monolayer-bilayer junction of hole-doped MoTe$_2$
at the electron density of $n=-7.02\times10^{13}\mathrm{cm}^{-2}$.
We first consider a situation where an incident electron comes from the monolayer region
with the initial angle $\theta_m$,
and transmitted to the bilayer region with the finial angle $\theta_b$,
as shown in Fig.\ref{fig_refract_mote2}(a).
Here we adopt the electron picture rather than the hole picture
in describing the carrier transmission
although the system is hole-doped.
The traveling angle is defined by
$\theta = \arctan(v_y/v_x)$ from the expectation value of the 
velocity $(v_x,v_y)$ of the corresponding electronic state.
In Fig.\ \ref{fig_refract_mote2}(b), the relation between the initial and final angles 
is plotted separately for up-spin and down-spin electrons.
The corresponding transmission probability is plotted
in Fig.\ \ref{fig_refract_mote2}(c) as a function of the incident angle $\theta_m$.
Since the system is time-reversal symmetric,
Fig.\ \ref{fig_refract_mote2}(b) can be inversely viewed
as the angle relationship (with spin inverted) for an electron coming from the bilayer region
with the initial angle $\theta_b$ and transmitted to monolayer region
with the final angle $\theta_m$.
Fig.\ \ref{fig_refract_mote2}(c) then represents the transmission probability
as a function of the final angle $\theta_m$.

Fig.\ \ref{fig_refract_mote2}(b) shows that
the atomic step is highly angle-selective for
an incident carrier from the monolayer side,
i.e. it allows to pass 
an up-spin electron only within a narrow range of angle $-30^\circ\,\lsim\,\theta_m\,\lsim\,-7^\circ$,
and down-spin electron only within $7^\circ\,\lsim\,\theta_m\,\lsim\,30^\circ$,
while the transmitted carriers widely spread in $-90^\circ<\theta_b<90^\circ$ in the bilayer region.
If an electron comes from the bilayer region, on the contrary,
any incident angles can be allowed 
while the transmitted electrons are highly collimated in the monolayer 
to the different angle ranges
depending on spins.


The spin-dependent refraction is attributed to the difference in the Fermi surface 
structure between the monolayer and bilayer TMDs, which are illlustrated 
in Fig.\ \ref{fig_refract_mote2}(d).
The monolayer's Fermi surface is completely spin-split,
and the up-spin and down-spin branches are located at $K'$ and $K$ points, respectively.
On the other hand,
the bilayer's Fermi surface is completely spin degenerate,
and also it has the third pocket at $\Gamma$ point besides $K$ and $K'$.
Due to the condition of a homogeneous carrier density,
the total area enclosed by the Fermi surfaces is exactly equal in monolayer and bilayer.
As a consequence, the size of the $K$ and $K'$ pockets are
significantly smaller in the bilayer than in the monolayer,
because in bilayer, the Fermi surface is doubled due to the spin degeneracy
and also the area of the $\Gamma$ pocket
reduces the share of $K$ and $K'$.

Because of the translation symmetry in the $y$ axis, 
the transverse momentum $p_y$ is preserved
in the transmission process.
The change of the traveling angle can be found
by comparing the velocity vectors (normal to the Fermi surface)
of the monolayer and bilayer states sitting at the same $p_y$.
As shown in Fig.\ \ref{fig_refract_mote2}(d),
the smaller Fermi surface of the bilayer 
restricts the corresponding monolayer region
to a portion of the whole Fermi surface,
in which the velocity direction varies only slightly
due to the trigonal warped character of the equi-energy surface.
This explains the reason why monolayer's electrons within only a small angle range
can transmit to the bilayer region.
As the up-spin ($K'$) and down-spin ($K$) Fermi surfaces 
are mirror symmetric about the $x$ axis, 
the up-spin and down-spin electrons are refracted in the opposite direction
with respect to $\theta=0$, as shown in Fig.\ \ref{fig_refract_mote2}(c).


\begin{figure}[htbp]
	\includegraphics[width=85mm]{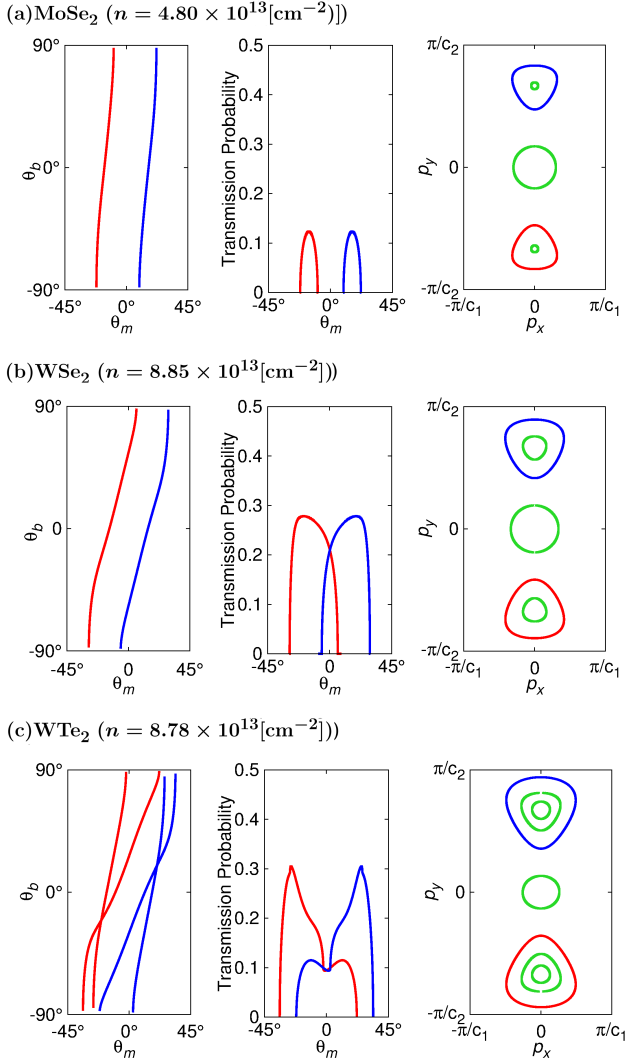}
	\caption{Similar plots to Fig.\ \ref{fig_refract_mote2} for (a)MoSe$_2$, (b)WSe$_2$, and (c)WTe$_2$. 
	}\label{fig_refract_others}
\end{figure}

\begin{figure}[htbp]
	\includegraphics[width=85mm]{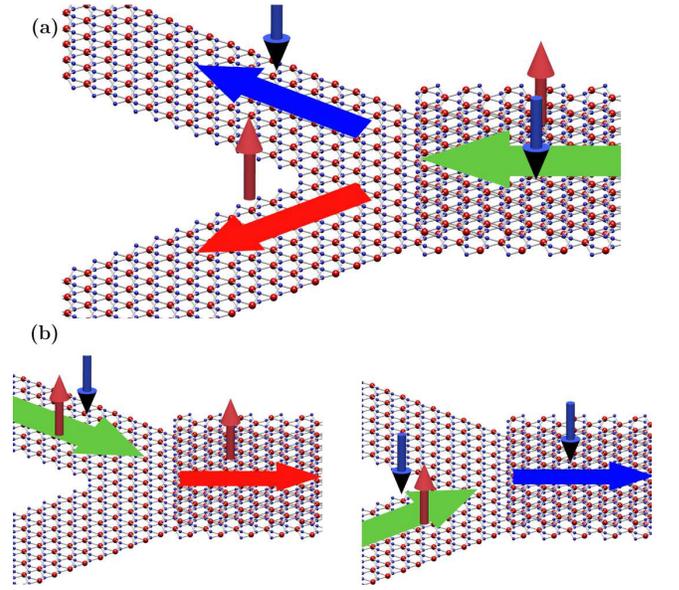}
	\caption{
		Spin filtering effect in Y-junction fabricated on the
		monolayer-bilayer boundary of hole-doped TMD,
		for the electron beam injected from (a) the bilayer region and (b) the monolayer region. 
	}\label{fig_y_junc}
\end{figure}



The carrier density assumed in Fig.\ \ref{fig_refract_mote2}
was chosen in such a way that the Fermi energy
comes slightly above the lower branch of the spin-split valence bands
of the monolayer MoTe$_2$.
This is the maximum hole density (i.e., the lowest Fermi energy) under the condition that 
each of $K$ and $K'$ valleys is dominated by a single spin. 
If we raise the Fermi energy toward the valence band maximum, 
the difference in the refraction angle between up and down spins
gradually decreases.
This is because the Fermi surface becomes more circular
near the band edge (i.e., the trigonal warping is weaker)
so that the transmission becomes more symmetric with respect to $\theta=0$,
and that reduces the spin dependence of the transmission.


Fig.\ \ref{fig_refract_others} presents similar calculations for 
other TMDs, MoSe$_2$, WSe$_2$, and WTe$_2$.
We see that those materials share a basically similar characteristics of the Fermi surface 
and the nature of the spin-dependent refraction.
In the transition-metal disulfides MS$_2$ (not shown), on the other hand, 
the valence electrons are completely reflected at the atomic step,
because the low-energy spectrum of the bilayer is dominated by $\Gamma$ point,
and the transmission from $K$ and $K'$ points of the monolayer side
is completely blocked due to the momentum mismatch.


We can exploit the atomic step on TMD as a spin splitter
which spatially separates up-spin and down-spin electrons,
simply by passing the electric current through the boundary.
For example, we can consider a Y-shaped junction fabricated on 
monolayer-bilayer boundary of hole-doped TMDs as shown in Fig.\ \ref{fig_y_junc},
where a single strip in the bilayer region forks into two branches in the monolayer region.
If the spin-unpolarized electric current is injected from the bilayer side,
it separates into the up-spin and down-spin currents in different branches of the monolayer
due to the spin-dependent collimation effect, as shown in Fig.\ \ref{fig_y_junc}(a).
When the spin-unpolarized current is put from 
one of the monolayer branches, on the contrary,
only a single spin can enter the bilayer region [Fig.\ \ref{fig_y_junc}(b)],
because of the spin and angle selective transmission from the monolayer region.


To conlcude, we studied the electron transmission at the monolayer-bilayer atomic step of TMDs,
and find the spin-dependent refraction effect
which separates the up-spin and down-spin carriers into different traveling directions.
The phenomena suggest a potential application for the spintronic devices
which transfers the spin information into the electric information.

We would like to thank Professor Takashi Koretsune for his help in using the quantum-ESPRESSO 
and Wannier90 packages.
This work was supported by Grants-in-Aid for Scientific research (GrantsNo. 24740193 and No. 25107005).
Wannier90: A Tool for Obtaining Maximally-Localised Wannier Functions
A. A. Mostofi, J. R. Yates, Y.-S. Lee, I. Souza, D. Vanderbilt and N. Marzari
Comput. Phys. Commun. 178, 685 (2008) [ONLINE JOURNAL]


\appendix
\section{Tight-binding model for the monolayer-bilayer junction}

\begin{figure}[htbp]
	\includegraphics[width=80mm]{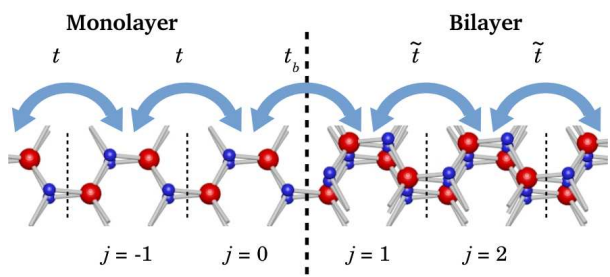}
	\caption{
		Tight-binding model for the TMD monolayer-bilayer junction.
		Red and blue spheres represent transition-metal and chalcogenide atoms, resprectively.
		Unit cells in the monolayer (bilayer) contain two (four) transition-metal atoms and four (eight) chalcogenide atoms.
	}\label{fig_app}
\end{figure}

Figure \ref{fig_app} schemtically illustrates the tight-binding model
for the TMD monolayer-bilayer junction.
The system is translationally symmetric in the $y$ direction,
so that it is reduced to a one-dimensional system labeled by the momentum $p_y$. 
The unit cell of monolayer (bilayer) region
contains two (four) transition-metal atom and four (eight) chalcogenide atoms.
Note that the unit cell here is twice as large as the primitive unit cell of the bulk TMD
because of the armchair geometry.
We consider five $d$-orbitals for each transition-metal atom 
and three $p$-orbitals for each chalcogenide atom,
which sum up to 22 (44) orbitals per spin in monolayer (bilayer) unit cell.
The hopping parameters of the tight-binding model
are derived by 
computing the maximally localized Wannier Functions by applying Wannier90
to the first principle electronic density.
First we construct the spin-less tight-binding model neglecting the spin-orbit interaction,
and introduce the spin-orbit coupling
by adding the appropriate $\textbf{L}\cdot\textbf{s}$ term to the transition-metal atoms. \cite{Liu2013}
The up-spin and down-spin states (with respect to $z$ axis)
are completely decoupled even in presence of the spin-orbit coupling,
and therefore we can calculate the transmission probability
for each spin state separately.

The tight-binding Hamiltonian for the monolayer-bilayer junction
for given $p_y$ and spin state $s=\uparrow, \downarrow$ is written as
\begin{align*}
	H(p_y,s)=&
	\sum_{j\leq0}\left[
	{\boldsymbol{c}_{j}}^\dagger h{\boldsymbol{c}_{j}}
	+
	({\boldsymbol{c}_{j}}^\dagger {t}^\dagger{\boldsymbol{c}_{j-1}}
	+
	\mathrm{h.c.})
	\right]\\
	&+
	\left(
	{\tilde{\boldsymbol{c}}_{1}} ^\dagger {t_b}^\dagger{\boldsymbol{c}_{0}}
	+
	\mathrm{h.c.}
	\right)\\
	&+
	\sum_{1\leq j}\left[
	{\tilde{\boldsymbol{c}}_{j}} ^\dagger \tilde{h}{\tilde{\boldsymbol{c}}_{j}}
	+
	({\tilde{\boldsymbol{c}}_{j+1}} ^\dagger {\tilde{t}}^\dagger{\tilde{\boldsymbol{c}}_{j}}
	+
	\mathrm{h.c.})
	\right].
\end{align*}
Here the index $j$ indicates the cell position,
where $j\leq 0$ and $j\geq 1$ 
correspond to the monolayer and bilayer regions, respectively.
The vector $\boldsymbol{c}_j \, (j\leq 0)$ is the $N(=22)$-component annihilation operator
of electron at the cell $j$ in the monolayer region, where each component correspond to the 
atomic orbital inside the unit cell.
Similarly, $\tilde{\boldsymbol{c}}_j \, (j\geq 1)$ 
the $2N$-component annihilation operator for the bilayer region.
$h$($\tilde{h}$) and $t$($\tilde{t}$) are $N\times N$($2N\times2N$) matrices,
where
$h$($\tilde{h}$) describes the matrix elements inside the unit cell
and $t$($\tilde{t}$) describes the matrix elements between neighboring cells
in the monolayer (bilayer) region.
The hopping between monolayer and bilayer regions (i.e., between $j=0$ and 1)
is described by $t_b$, which is $N\times2N$ matrix.
In this paper, we assume that $t_b$ is equivalent to the half submatrix of $\tilde{t}$,
i.e., we borrow all the hopping parameters from bulk bilayer
and neglect the non-existing upper layer in the monolayer side.
We can safely neglect further hopping elements between $j$ and $j+n\, (n\geq 2)$ ,
which are sufficiently small.

\bibliography{TMD}

\end{document}